\documentclass[twocolumn,aps,prb,showpacs,amsmath,superscriptaddress,longbibliography,notitlepage]{revtex4-1}
\usepackage{amssymb}
\usepackage{mathrsfs}
\usepackage{graphicx}
\usepackage{float}
\usepackage[caption=false]{subfig}
\usepackage[pdftex,colorlinks=red]{hyperref}

\usepackage{epstopdf}

\usepackage[normalem]{ulem}
\usepackage{verbatim}
\usepackage{xcolor}
\usepackage{bm}
\begin{document}

\title{Extrinsic higher-order topological corner states in AB-stacked transition-metal dichalcogenides}

\author{Jiang Yao}
\affiliation{Guangdong Provincial Key Laboratory of Quantum Metrology and Sensing $\&$ School of Physics and Astronomy, Sun Yat-Sen University (Zhuhai Campus), Zhuhai 519082, China}
\author{Linhu Li}\email{lilh56@mail.sysu.edu.cn}
\affiliation{Guangdong Provincial Key Laboratory of Quantum Metrology and Sensing $\&$ School of Physics and Astronomy, Sun Yat-Sen University (Zhuhai Campus), Zhuhai 519082, China}

\date{\today}

\begin{abstract}
    Higher-order topological insulators (HOTIs) are a novel type of topological phases which supports $d$-dimensional topological boundary states in D-dimensional systems with $D-d>1$.
    In this work, we theoretically predict that interlayer couplings in AB-stacked bilayer transition-metal dichalcogenides (TMDs) lead to the emergence of extrinsic second-order topological phases, 
    where corner states are induced by the band inversion of zigzag edge bands.
    We find that the systems feature a quantized multiband Berry phase defined for a zigzag nanoribbon geometry,
    unveiling the nontrivial topological properties of its two zigzag edges.
    With detailed investigation into 
    the bilayer TMDs under different geometries, we find two types of boundary-obstructed corner states arising from different corner terminations of either the same type of or heterogeneous zigzag edges.
    The topological nature of these corner states and their degeneracy is further analyzed with both the crystalline symmetries of different geometries, and a topological phase transition of the Berry phase induced by a layer-dependent onsite energy.
    \end{abstract}

\maketitle

\section{Introduction}
Topological insulators are materials that exhibit robust boundary states, protected by a band gap and distinctive topological properties of their bulk states~\cite{hasan2010colloquium,qi2011topological}.
%These materials can be classified by different symmetries and ``orders" of topology, namely, a 
Their topological  features can be classified into different ``orders", namely,
a $n$th-order topological insulators in $d$-dimension (dD) host topological boundary states in its $(d-n)$D boundaries.
%~\cite{ryu2010topological,chiu2016classification,li2018direct,geier2018second,trifunovic2019higher,khalaf2018higher,okuma2019topological,rasmussen2020classification,li2021direct,lei2022topological}.
Topological insulators with their orders of topology higher than $1$ are known as the higher-order topological insulators(HOTIs)~\cite{benalcazar_electric_2017,benalcazar_quantized_2017}, 
whose realization has been proposed and implemented in both materials~\cite{schindler_higher-order_2018,schindler_higher-order_2018_1,park_higher-order_2019,fang_higher-order_2020,agarwala_higher-order_2020} and quantum simulation setups
such as two-dimensional dielectric photonic crystals~\cite{xie_visualization_2019}, acoustic crystals ~\cite{wei_higher-order_2021,xue_acoustic_2019,ni_observation_2019,yang_helical_2020} and two-dimensional continuous elastic system~\cite{fan_elastic_2019}.

%Topological insulator are materials with gapped band structure characterized by quantized topological invariants that defined with respect to symmetries of their bulk Hamiltonian. 
%Different topological invariants correspond to different number of edge states in open boundary condition. This corresponding relations is called Bulk-Boundary Corresponding. 
%In a d-dimensional (dD) topological insulator, a topologically nontrivial bulk band structure implies the existence of (d-1)D boundary states, characterizing by nontrivial topological invariant defined with bulk band.
%However, Higher-order topological insulator (HOTIs) exhibits lower dimensional topological states. In 2017, the concept of higher order topological insulators is introduced and characterized by quantized multipole \cite{benalcazar_electric_2017,benalcazar_quantized_2017}.
%Higher-order topological phase is divided into ``intrinsic'' and ``extrinsic''. ``intrinsic'' higher-order topological phase depends on the crystalline symmetry, extrinsic higher-order topological phase does not depend on the crystalline symmetry, but on boundaries.
%And extrinsic higher-order topological phase also kown as boundary obstructed topological phase(BOTP),which rather than close bulk gap, the boundary gap close when a topological phase transition occurs \cite{geier_second-order_2018,khalaf_boundary-obstructed_2021}.

Over the past decade,
exploring novel topological phases such as HOTIs in natural electronic systems has been one of the most active research topics in condensed matter physics and material science~\cite{xie_higher-order_2021,bernevig2022progress,wieder2022topological}. Among the variety of natural materials,
semiconducting 2D transition metal dichalcogenides (TMDs) possess a large bulk gap and in-gap 1D boundary states~\cite{rostami_edge_2016},
thus they
provide an ideal platform for investigating various boundary phenomena.
%such as edge magnetism~\cite{brito2022edge}, 
%Majorana edge modes~\cite{chu2014spin,xu2014topological,li_strain-induced_2016,deng2019superconducting}, 
%and confined 1D states at the twin-grain-boundary defect~\cite{van2013grains,liu2014dense,barja2016charge,cadez_robust_2019}.
In particular, HOTI phases have been demonstrated to
arise from double band inversion of surface states in the $\beta$- and $\gamma$- phases of TMDs~\cite{wang2019higher},
or from 
%Kagome lattice structure 
staggered coupling amplitudes between different orbitals
and $C_3$ rotation symmetry 
in TMD monolayers~\cite{zeng_multiorbital_2021,qian_c_2022,jung_hidden_2022},
whose connection to the orbital Hall effect has been recently unveiled~\cite{costa2023connecting}.

% inhabit the $\beta$- and $\gamma$- phases of TMDs due to double band inversion of surface states~\cite{wang2019higher},
%or TMD monolayers protected by $C_3$ rotation symmetry~\cite{zeng_multiorbital_2021,qian_c_2022,jung_hidden_2022}, whose connection to the orbital Hall effect has been recently unveiled~\cite{costa2023connecting}.

In this paper, we predict the presence of higher-order topological corner states in AB-stacked bilayer TMDs, 
based on a three-band tight-binding model describing the low-energy and edge-state physics in monolayers of group-VIB TMDs\cite{liu_three-band_2013}.
Recently, such structures have been shown to support different types of first-order topological phases~\cite{pan2022topological}.
Conventionally, corner states are expected to be more accessible in TMDs with armchair boundaries, which possess a large energy gap between the 1D boundary states, whilst zigzag boundary states are gapless and may overwhelm possible corner states~\cite{rostami_edge_2016,zeng_multiorbital_2021,qian_c_2022}.
However, 
we find that interlayer couplings can induce a zigzag-boundary band inversion for bilayer TMDs,
and generates corner states corresponding to the boundary obstructed topological phases with ``extrinsic" higher-order topology, without relying on crystalline symmetries of the system\cite{geier2018second,trifunovic2019higher,ezawa2020edge,asaga2020boundary,wu2020boundary,tiwari2020chiral,khalaf_boundary-obstructed_2021,li2018direct,li2021direct,lei2022topological}.
Unlike a single monolayer,
AB-stacking structure allows for two types of spatially symmetric zigzag boundaries, formed by boundaries of the two layers with different atoms.
Interestingly, we find that different terminations between them give arise to two classes of corner states,
where only one of them shows a direct correspondence to boundary-gap closing and a topological transition characterized by a Berry phase.
These rich phenomena are exhaustively investigated with triangular, hexagonal, and parallelogram geometries, which support either one or both types of zigzag boundaries and thus different corner terminations.

%In previous studies\cite{zeng_multiorbital_2021,qian_c_2022,jung_hidden_2022}, TMDs as higher-order topological crystalline insulators(HOTCIs) with corner states was studied. Higher-order topological corner states in their work is sensitive to bulk crystalline symmetry.
%Our work predicts the presence of extrinsic higher-order topological corner states in AB-stacked TMDs with many structures, which is crystalline symmetry independent and boundary dependent.
%Corner states induced by boundary band inversion, which is gapless zigzag band gapped by interlayer hopping in AB-stacked TMDs. Besdies, the corner states can be countrolling by tunning on-site energy in different layers. 

%In the following, all calculation are based on parameters 
%which from the first-principles calculation with the generalized gradient approximation for $\text{MoS}_2$ in Ref \cite{liu_three-band_2013}.
%Unless otherwise stated, $\text{MoS}_2$ refers to TMDs.

The rest of this paper is organized as follows.
In Sec.\ref{model}, we introduce the three-orbital tight-binding model we use to describe AB-stacked bilayer TMDs,
including their nontrivial topology characterized by a Berry phase,
and crystalline symmetries that assist our analysis of corner states under different geometries.
%And then introduce the crystalline symmetry of bulk state, different boundaries in AB-stacked TMDs, demonstrate band inversion in zigzag band finally.
In Sec. \ref{sec_confi}, we study the emergence of corner states and their behaviors during topological phase transitions of the Berry phase, in triangular, hexagonal, and parallelogram lattices respectively, and analyze their spatial configurations with the help of presence and absence of different crystalline symmetries.
A summary and some discussion of our results are given in Sec. \ref{Conclusion}.

\section{model}\label{model}
\subsection{Lattice structure and Hamiltonian}
\begin{figure*}
    \includegraphics[width=\linewidth]{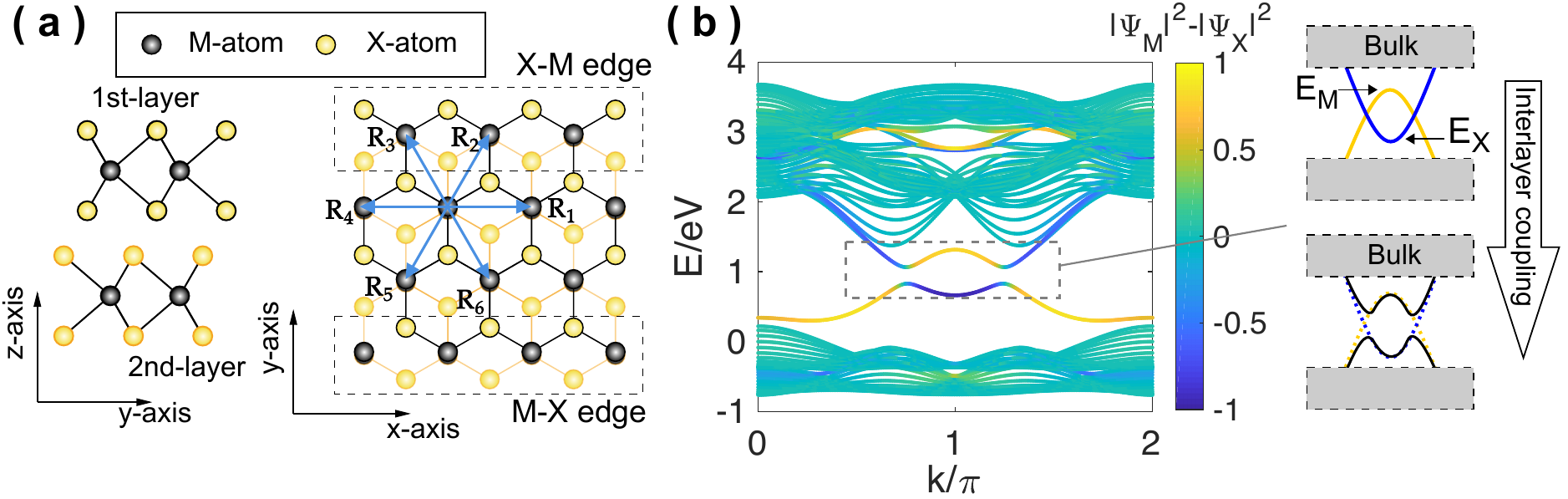}
    \caption{Lattice structure and energy spectrum of AB-stacked bilayer TMDs.
    (a) Side view and top view of the model, with $\bm{R}_1$ to $\bm{R}_6$ indicating the nearest neighbors. 
    Dashed boxes show different zigzag edges in AB-stacked TMDs.
    (b) Energy spectrum of a zigzag nanoribbon, with $\nu=0.3$ eV and $\mu=0$.
    Eigenstates are marked by different colors according to the quantity $|\Psi_{\rm M}|^2-|\Psi_{\rm M}|^2$,
    where $|\Psi_M|^2$ ($|\Psi_X|^2$) is the sum distribution at M edges (X edges) of the two layers for each eigenstate.
    Namely, $|\Psi_{\rm M}|^2-|\Psi_{\rm M}|^2\approx 1$ for M edge states, and $\approx -1$ for X edge states.
    Right panel in (b) sketches the band inversion mechanism of zigzag edge bands.
    $E_{\rm M}$ and $E_{\rm X}$ are the maximal value of M-edge band and the minimal value of X-edge band, respectively. 
    For $\text{MoS}_2$, $E_{\rm M}\approx 1.32$ eV and $E_{\rm X} \approx 0.65$ eV.}
    \label{fig1_model}
\end{figure*}

The systems we consider are AB-stacked bilayer TMDs, 
commonly referred as MX$_2$ with M and X denoting atoms of transition metals and chalcogens respectively, 
with a honeycomb-like structure as sketched in Fig. \ref{fig1_model}(a).
To analyze the edge and corner physics we concern,
we adopt a three-orbital model for each layer~\cite{liu_three-band_2013}, constructed using only the $d_{z^2}$, $d_{xy}$, and $d_{x^2-y^2}$ orbitals of M atoms,
which gives a reasonable description of zigzag edge states of these materials, and has been applied to investigate various edge phenomena therein~\cite{costa2023connecting,li_strain-induced_2016,cadez_robust_2019,brito2022edge}.
The tight-binding  Hamiltonian can be written as
\begin{equation}    \label{Hamiltonian}
    \hat{H}=\hat{H}_{\rm 1st}+\hat{H}_{\rm 2nd}+\hat{H}_{\rm int}+\frac{\mu}{2}(\hat{N}_1-\hat{N}_2),
\end{equation}
with $\hat{H}_{\rm 1st}$ ($\hat{H}_{\rm 2nd}$) the monolayer Hamiltonian for the top (bottom) layer,
$\hat{H}_{\rm int}$ the interlayer couplings,
and $\hat{N}_{1,2}$ the total electron number operator.
The last term of $\mu$ describes the difference of on-site energy for the two layers, which may be induced by external electric fields applied perpendicular to the layers~\cite{rama2011tubable,kumar2013semiconductor}.
Note that a nonzero $\mu$ is not essential to induce corner states in our systems. However, it can lead to a topological phase transition that changes the number of corner states, as demonstrated in later sections.
%the three-orbital tight-binding Hamiltonian described below as the first layer, $H_{2nd}$ can be obtain by inversing the hopping strenth  in real space in $H_{1st}$, and as the second layer, labelling different degrees of freedom, $H_{int}$ is the interlayer coupling dominated by $d_{z^2}$-$d_{z^2}$ orbitals'  coupling. $N_1$ and $N_2$ are particles number operator of different layers,respectively. $\mu$ is the difference in the on-site energy of the different layers~\LLH{[LLH: mention possible origin (electric field?) and cite some papers]}.

Explicitly, the first two Hamiltonian operators are given by
\begin{eqnarray}
        \hat{H}_{\rm 1st}&=&\sum_{i,\bm{R}}\sum_{\alpha,\alpha'}\hat{a}_{i,\alpha}^\dagger 
    t_{\bm{R},\alpha,\alpha'} \hat{a}_{i+\bm{R},\alpha'} + \sum_{i,\alpha}\hat{a}_{i,\alpha}^\dagger  \epsilon_{\alpha} \hat{a}_{i,\alpha},\\
        \hat{H}_{\rm 2nd}&=&\sum_{i,\bm{R}}\sum_{\alpha,\alpha'}\hat{b}_{i,\alpha}^\dagger 
t_{\bar{\bm{R}},\alpha,\alpha'} \hat{b}_{i+\bm{R},\alpha'} + \sum_{i,\alpha}\hat{b}_{i,\alpha}^\dagger  \epsilon_{\alpha} \hat{b}_{i,\alpha},
\end{eqnarray}
Where $\hat{a}_{i,\alpha}^\dagger$ ($\hat{b}_{i,\alpha}^\dagger$) creates an electron at lattice site $i$ and orbital $\alpha$ in the top (bottom) layer,
$\bm{R}$ is one of the six vectors connecting nearest-neighbor M atoms with $t_{\bm{R},\alpha,\alpha'}$ being its corresponding hopping strength [see Fig. \ref{fig1_model}(a) and Table.~\ref{hopping}],
and $\epsilon_{\alpha}$ are on-site energies corresponding to different orbitals,

The third term $\hat{H}_{\rm int}$ in Eq.\ref{Hamiltonian} is the interlayer hopping. Using Slater-Koster table \cite{slater_simplified_1954}, $\hat{H}_{\rm int}$ reads
\begin{eqnarray}
        \hat{H}_{\rm int}&=&\hat{a}_{i,d_{z^2}}^\dagger V_{dd\sigma} \hat{b}_{i,d_{z^2}} + \hat{a}_{i,d_{xy}}^\dagger V_{dd\delta} \hat{b}_{i,d_{xy}}\nonumber\\
        &&+\hat{a}_{i,d_{x^2-y^2}}^\dagger V_{dd\delta} \hat{b}_{i,d_{x^2-y^2}}+H.c.,
\end{eqnarray}
Where $V_{dd\sigma}$ and $V_{dd\delta}$ are two types of overlap intergrals of different d-d orbitals,
and $V_{dd\sigma}$ is generally much larger than $V_{dd\delta}$ for a relatively large distance between M atoms \cite{jaffe_overlap_1953,roberts_overlap_1957}. 
In the following discussion, we shall assume 
\begin{eqnarray}
V_{dd\sigma}=\nu,~~V_{dd\delta}=0.3\nu,\label{eq:assumption}
\end{eqnarray}
and use the tight-binding parameters for MoS$_2$ given in Ref. \cite{liu_three-band_2013} ($\epsilon_{d_{z^2}}=1.046$ eV, $\epsilon_{d_{xy}}=\epsilon_{d_{x^2-y^2}}=2.104$ eV, ; see Table \ref{hopping} for other parameters), 
%to unveil the topological properties in our system, 
unless otherwise specified.
%\red{It is worth noting that interlayer hopping strength $\nu$ as the description of interlayer Van der Waals force, which is weaker than in-plane hopping, taking an absolute value of $0.75$ eV is difficult to exist steadily in real materials for AB-stacked TMDs.}
\begin{table*}
    \centering
    \caption{Hopping amplitudes $t_{\bm{R}, \alpha,\alpha^{\prime}}$ for TMDs~\cite{liu_three-band_2013,li_strain-induced_2016}.
    Different rows show hoppings between different orbitals, and different columns are for different spatial hopping vectors $\bm{R}$ for the top layer in Fig.\ref{fig1_model}. For the bottom layer, $\bm{R}_1$, $\bm{R}_2$, and $\bm{R}_3$ are exchanged with $\bm{R}_4$, $\bm{R}_5$, and $\bm{R}_6$, respectively.
    For MoS$_2$, the parameters are approximately given by $t_0=-0.184$, $t_1=0.401$, $t_2$=0.507, $t_{11}=0.218$, $t_{12}=0.338$, $t_{22}=0.057$ (in eV)~\cite{li_strain-induced_2016}.}
    \begin{tabular}{lcccccc}
        \hline \hline
       $\alpha\text{-}\alpha^{\prime}$ & $\bm{R}_1$ & $\bm{R}_2$ & $\bm{R}_3$ & $\bm{R}_4$ & $\bm{R}_5$ & $\bm{R}_6$ \\
        \hline
        $d_{z^2}\text{-}d_{z^2}$ & $t_0$ & $t_0$ & $t_0$ & $t_0$ & $t_0$ & $t_0$ \\
        $d_{x y}\text{-}d_{x y}$ & $t_{11}$ & $\frac{t_{11}+3 t_{22}}{4}$ & $\frac{t_{11}+3 t_{22}}{4}$ & $t_{11}$ & $\frac{t_{11}+3 t_{22}}{4}$ & $\frac{t_{11}+3 t_{22}}{4}$ \\
        \begin{comment}\makecell\end{comment}{$d_{x^2-y^2}\text{-}d_{x^2-y^2}$} & $ t_{22}$ & $\frac{3 t_{11}+t_{22}}{4}$ & $\frac{3 t_{11}+t_{22}}{4}$ & $t_{22}$ & $\frac{3 t_{11}+t_{22}}{4}$ & $\frac{3 t_{11}+t_{22}}{4}$ \\
        $d_{z^2}\text{-}d_{x y}$ & $t_1$ & $\frac{t_1+\sqrt{3} t_2}{2}$ & $-\frac{t_1+\sqrt{3} t_2}{2}$ & $-t_1$ & $-\frac{t_1-\sqrt{3} t_2}{2}$ & $\frac{t_1-\sqrt{3} t_2}{2}$ \\
        $d_{z^2}\text{-}d_{x^2-y^2}$ & $t_{2}$ & $-\frac{t_2-\sqrt{3} t_1}{2}$ & $-\frac{t_2-\sqrt{3} t_1}{2}$ & $t_{2}$ & $-\frac{t_2+\sqrt{3} t_1}{2}$  & $-\frac{t_2+\sqrt{3} t_1}{2}$ \\
        $d_{x y}\text{-}d_{x^2-y^2}$ & $t_{12}$ &  {$\frac{\sqrt{3}(t_{11}-t_{22})}{4}-t_{12}$} 
        &{$\frac{\sqrt{3}(t_{22}-t_{11})}{4}+t_{12}$} &$-t_{12}$ 
        &{$\frac{\sqrt{3}(t_{11}-t_{22})}{4}+t_{12}$}  & {$\frac{\sqrt{3}(t_{22}-t_{11})}{4}-t_{12}$} \\
        \hline \hline
        \end{tabular}
        \label{hopping}
\end{table*}

\subsection{Zigzag edge states and edge-band inversion in a nanoribbon structure}
\begin{figure}
    \includegraphics[width=\linewidth]{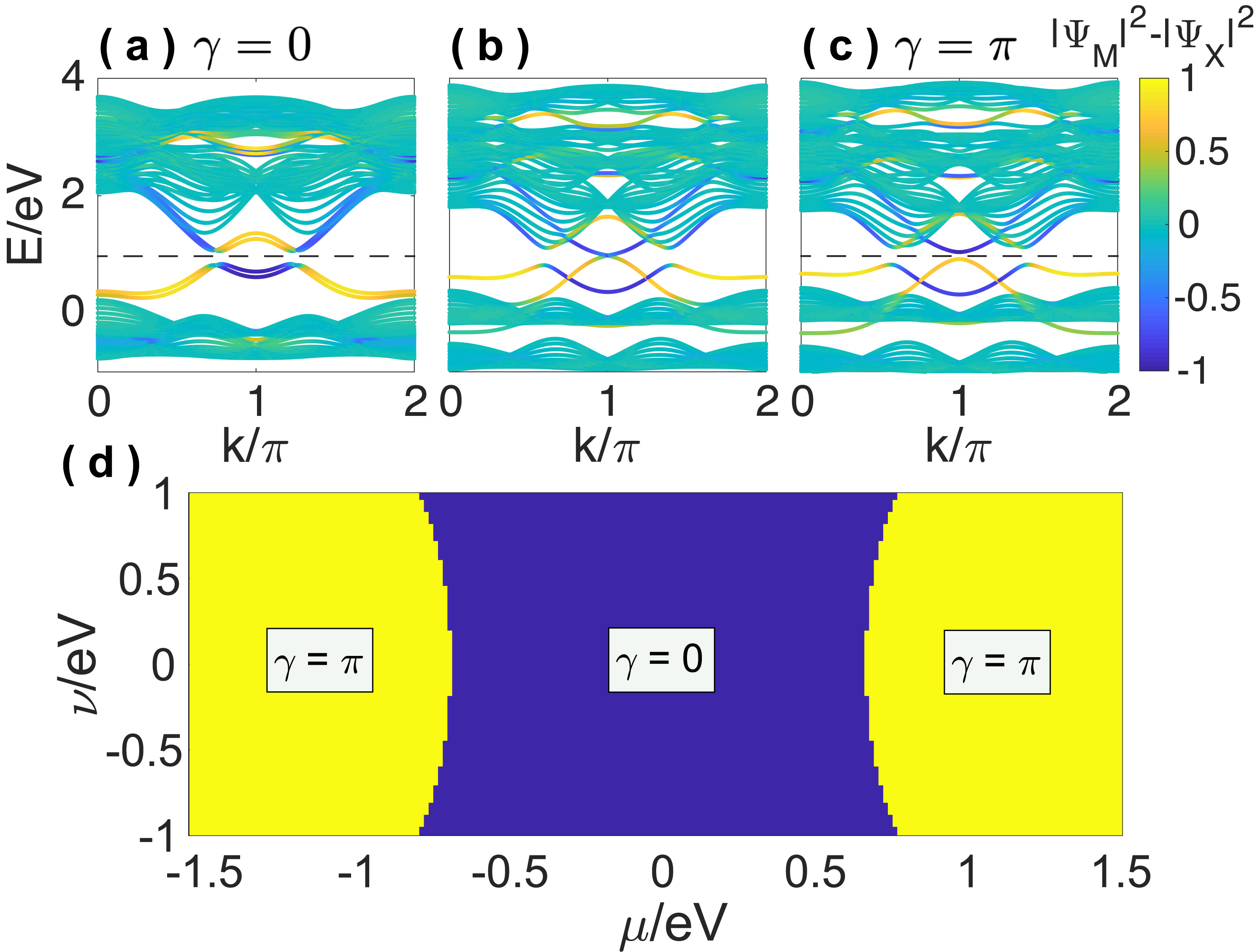}
    \caption{
    Energy spectrum and a topological phase diagram of the bilayer TMDs in a nanoribbon structure, with $\mu=0.1$ eV, $0.68$ eV, and $0.8$ eV from (a) to (c) respectively, and the same colormap as in Fig. \ref{fig1_model}, indicating distribution of each eigenstate at M and X edges.
    With increasing $\mu$, a reversed process of band inversion occurs for one pair of edge states, which becomes topologically trivial after the transition at $\mu\approx 0.67$ eV.
    In (a) and (c), $\gamma$ is calculated for all eigenstates below the band gap (indicated by the dash lines).
    We find $\gamma=0$ in (a) as edge-band inversion occurs for both M-X  and X-M edge states, which possess opposite single-band Berry phase (see Appendix \ref{Berry phase}).
    In (c), we have $\gamma=\pi$ since one pair of these edge states is trivialized after the topological phase transition.
    1D edge states in (b) are gapless, and thus the Berry phase is ill-defined. 
    Other parameters are the same as in Fig. \ref{fig1_model}.
    (d) A phase diagram determined by the value of $\gamma$.}
    \label{fig:Evsmu}
\end{figure}

To reveal the topological nature that gives raise to higher-order corner states,
we take a look at the system with a zigzag nanoribbon structure along $x$ direction.
When $\mu=\nu=0$, our model reduces to two identical monolayers of TMDs, 
where edge bands of different zigzag edge states (namely, M-edge and X-edge) of the two layer cross each other in their eigenenergies. 
The AB-stacked structure allows for a mixture of the two branches of edge states on the same edge of the bilayer TMD, denoted as M-X  or X-M edge according to the boundary atoms of the two layers [see Fig. \ref{fig1_model}(a)].
Thus, with a nonzero $\nu$ that couples M- and X-edges of different layers, 
an edge-band inversion occurs and opens a boundary gap between these edge states,
as shown in Fig. \ref{fig1_model}(b) and (c).
Note that these edge states are two-fold degenerate when $\mu=0$, due to an inversion symmetry between M-X  and X-M edges.
With a nonzero $\mu$, the degeneracy is lifted, 
and one pair of edge states is separated in energy (the one at X-M edge for a positive $\mu$), resulting in a reversed process of band inversion and thus a topological phase transition,
as shown in Fig. \ref{fig:Evsmu}. These different topological phases can be further characterized by a multi-band Berry phase $\gamma$, defined as
\begin{equation}
    \gamma=- i \sum_l \log \det U(k_l),\label{eq_BP}
\end{equation}
where $U_{mn}(k_l)=\langle{\psi_m(k_l)}|{\psi_n(k_{l+1})}\rangle$ is the $(m,n)$ element of
the link matrix $U(k_l)$, 
$|\psi_m(k_l)\rangle$ is the Bloch wavefunction of the $m$-th band at the discrete crystal momentum $k_l$,
and $m,n\in [1,N_{\rm occ} ]$ with $N_{\rm occ}$ the number of occupied bands (all bands below the edge-gap in our case, see Fig. \ref{fig:Evsmu}).

Numerically, we find that $\gamma=0$ for small $\mu$ with band inversion for both M-X  and X-M edge states [Fig. \ref{fig:Evsmu}(a)],
 as they possess opposite topological charges characterized by single-band Berry phases of $\pm\pi$ respectively (see Appendix \ref{Berry phase}).
Nonetheless, corner states may still emerge under the full OBCs with different geometries, as demonstrated in later sections. This is because the 1D edges are spatially separated by the bulk, therefore their opposite topological charges cannot annihilate each other.
On the other hand, increasing $\mu$ will lead to a topological phase transition and trivialize one pair of edge states, resulting in a $\pi$ Berry phase contributed solely by the other pair with edge-band inversion, as illustrated in Fig. \ref{fig:Evsmu}(c).
Following this analysis, the topological phase transition shall occur when the amplitude of $\mu$ matches the energy difference between the maximal value of M-edge band and minimal value of X-edge band [see Fig. \ref{fig1_model}(b)], $\mu\approx |E_M-E_X|\approx 0.67$ eV,
which is verified in the explicit examples discussed in later sections.
A topological phase diagram regarding different values of $\nu$ and $\mu$ is displayed in Fig. \ref{fig:Evsmu}(d), which shows that the strength of interlayer coupling $\nu$ does not affect much the Berry phase or the topological transition induced by $\mu$.
In other words, the nontrivial topology and its corresponding higher-order corner states are not sensitive to the exact value of interlayer couplings, and thus can be expected to manifest in a more realistic parameter regime [compared with our assumption in Eq. \eqref{eq:assumption}] of TMD materials.

\subsection{Crystalline symmetries of the AB-stacked bilayer TMDs}\label{sec_sym}
Although no essential to the extrinsic higher-order corner states, the lattice structure of TMDs naturally hosts several crystal symmetries, which are useful in our analysis of corner states under different geometries.
In particular, The space symmetry group of
monolayer TMDs is the $D_{3h}$ point group, containing symmetry operators
$\left\{ \hat{E},\hat{C}_3,\hat{C}_3^2,\hat{M}_x,\hat{M}_1,\hat{M}_2\right\}$,
where $\hat{E}$ is the identity operation,$\hat{C}_3$ is the rotation by $\frac{2\pi}{3}$ about the $z$-axis, 
$\hat{M}_x$ is the mirror-reflection along $x$-axis,
and $\hat{M}_1$ and $\hat{M}_2$
are obtained through rotating $\hat{M}_x$ around the $z$-axis by $\frac{2\pi}{3}$ and $\frac{4\pi}{3}$, respectively.
In addition to these symmetries,
AB-stacked bilayer TMDs (without a boundary) further satisfy a 3D inversion symmetry described by the operator $\hat{I}=\sigma_x \hat{\bar{ R}}$, with $\sigma_x$ exchanges the two layers, and $\hat{\bar{ R}}$ the central rotation around $z$-axis with 180 degree.
An extra interlayer-mirror symmetry along $y$-axis also emerges, described by the operator $\hat{M}'_y=\hat{I} \hat{M}_x$,
which represents a combination of mirror-reflection along $y$-axis and exchanging the two layers.
This symmetry is also equivalent to a C$_2$ symmetry around $x$ axis in 3D.

\section{Corner states under different geometric structures}\label{sec_confi}
As seen in Fig. \ref{fig:Evsmu}, nontrivial topological properties in our model originate from an edge-band inversion, which shall lead to the emergence of boundary obstructed topological corner states in the edge-gap near $E=0.95$ eV [dash lines in Fig. \ref{fig:Evsmu}] under the full OBCs.
More intriguingly, these corner states may show distinguished behaviors depending on the geometry of a OBC lattice,
since shearing the lattice along different directions will result in junctions either between heterogeneous M-X  and X-M edges, or two edges of the same type.

%Hence there are two types of boundaries in AB-stacked TMDs: boundary connected to homogeneous ``M-X edge'' and boundary connected to heterogeneous ``M-X edge''.
%For simplified, triagular structure with homogeneous ``M-X edge'' and hexagonal strucutre with heterogeneous ``M-X edge'' alternately are discussed then.

\subsection{Triangular structure}\label{Triangle}
\begin{figure}
    \includegraphics[width=\linewidth]{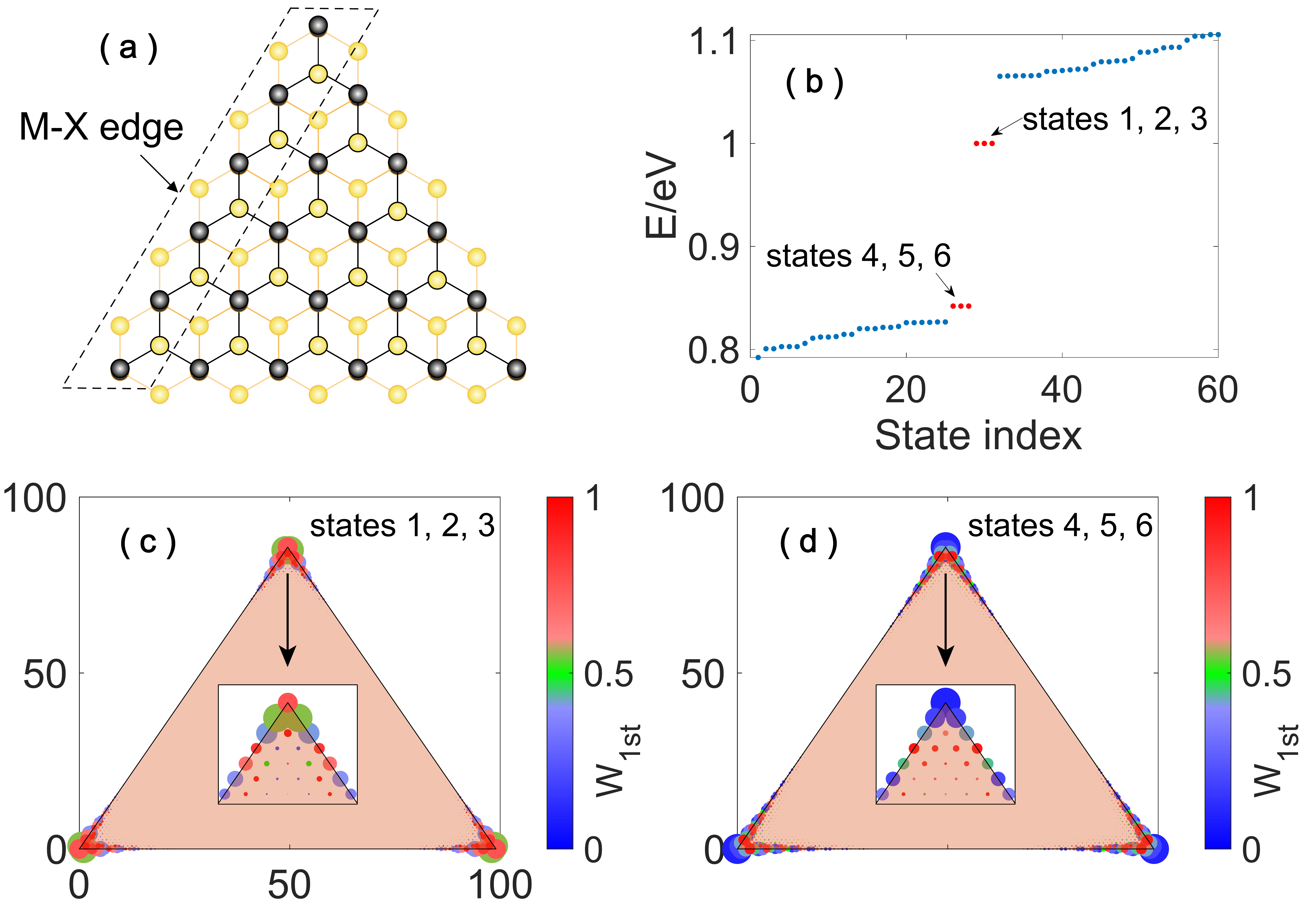}
    \caption{Lattice structure and corner states in AB-stacked bilayer TMDs with a triangular structure. 
    (a) Top view of the triangular lattice.
    (b) $60$ Eigenenergies close to $E=0.95$ eV, with two sets of three-fold degenerate corner states (red) within the energy gap.
    (c) and (d) distribution of the two sets of three-fold degenerate corner states respectively.
    The size of each point is proportional to the summed distribution at each site $i$, $\rho_{i}=\sum_{\alpha,n}(|\psi_{i,\alpha,n}^{\rm 1st}|^2+|\psi_{i,\alpha,n}^{\rm 2nd}|^2)$, with $\alpha$ denoting the three orbitals and $n$ summing over the three states indicated in each panel.
    The colormap displays the value of $W_{i}^{\rm 1st}$, the weight of occupation on the first layer for each site. Namely, $W_{i}^{\rm 1st}=1$ ($0$) means that the states occupy only the first (second) layer on the lattice site indexed by $i$.
    The system's size is chosen to have $100$ M atoms along each edge. Other parameters are $\nu=0.3$ eV and $\mu=0$.}
    \label{fig3_tri}
\end{figure}
We first consider AB-stacked bilayer TMDs with a triangular geometry, as shown in Figure.\ref{fig3_tri}(a). 
A triangular lattice has three equivalent boundaries of the same type, and here we take the case with M-X edges as an example.
%Such a geometric structure has three equivalent boundaries  which is named as ``M-X edge".
 In addition, it satisfies the 3-fold rotation symmetry of $\hat{C}_3$, but not the inversion symmetry that exchanges the two layers.
 Thus the corner states in a triangular bilayer TMDs are expected to be symmetric between different corners, and asymmetric between the two layers.
 %In addition another,  symmetric operators of crystal operators constitute a group:
%$\{ \hat{E}, \hat{C}_3,\hat{C}_3^2 ,\hat{M}_1,\hat{M}_2,\hat{M}_x \}$. Inersion symmetry is broken, meanwhile the properties of boundary are describes by either ``M-X edge'' band or ``X-M edge'' band.

In Fig. \ref{fig3_tri}(b), we display our numerically results of eigenenergies for the triangular structure. 
Two groups of three-fold degenerate states with different energies are found within the edge-gap.
Their distributions in real space are shown in Fig. \ref{fig3_tri}(c) and (d), where each set of degenerate states distributes symmetrically on the three corners. On the other hand, the two sets of corner states exhibit different occupation on the two layers, as can be seen from the weight of occupation on each site of the first layer,
$$W_{i}^{\rm 1st}=\sum_{n,\alpha}|\psi_{i,\alpha,n}^{\rm 1st}|^2/\left(\sum_{n,\alpha}|\psi_{i,\alpha,n}^{\rm 1st}|^2+\sum_{n,\alpha}|\psi_{i,\alpha,n}^{\rm 2nd}|^2\right),$$
with $\psi_{i,\alpha,n}^{\rm 1st (2nd)}$ the amplitude at the $i$th site of the 1st (2nd) layer of an eigenstate indexed by $n$, and the summation runs over all ortbiarls (indexed by $\alpha$) for a set of degenerate corner states.
Namely, states 1, 2, and 3 show a roughly balanced occupation on the two layers, and states 4, 5, and 6 mostly occupy the 2nd layer.
Such asymmetric layer occupation for corner states at different eigenergies reflects the absence of the inversion symmetry of $\hat{I}=\sigma_x\hat{\bar{R}}$ in the triangular lattice.
\begin{figure}
    \includegraphics[width=\linewidth]{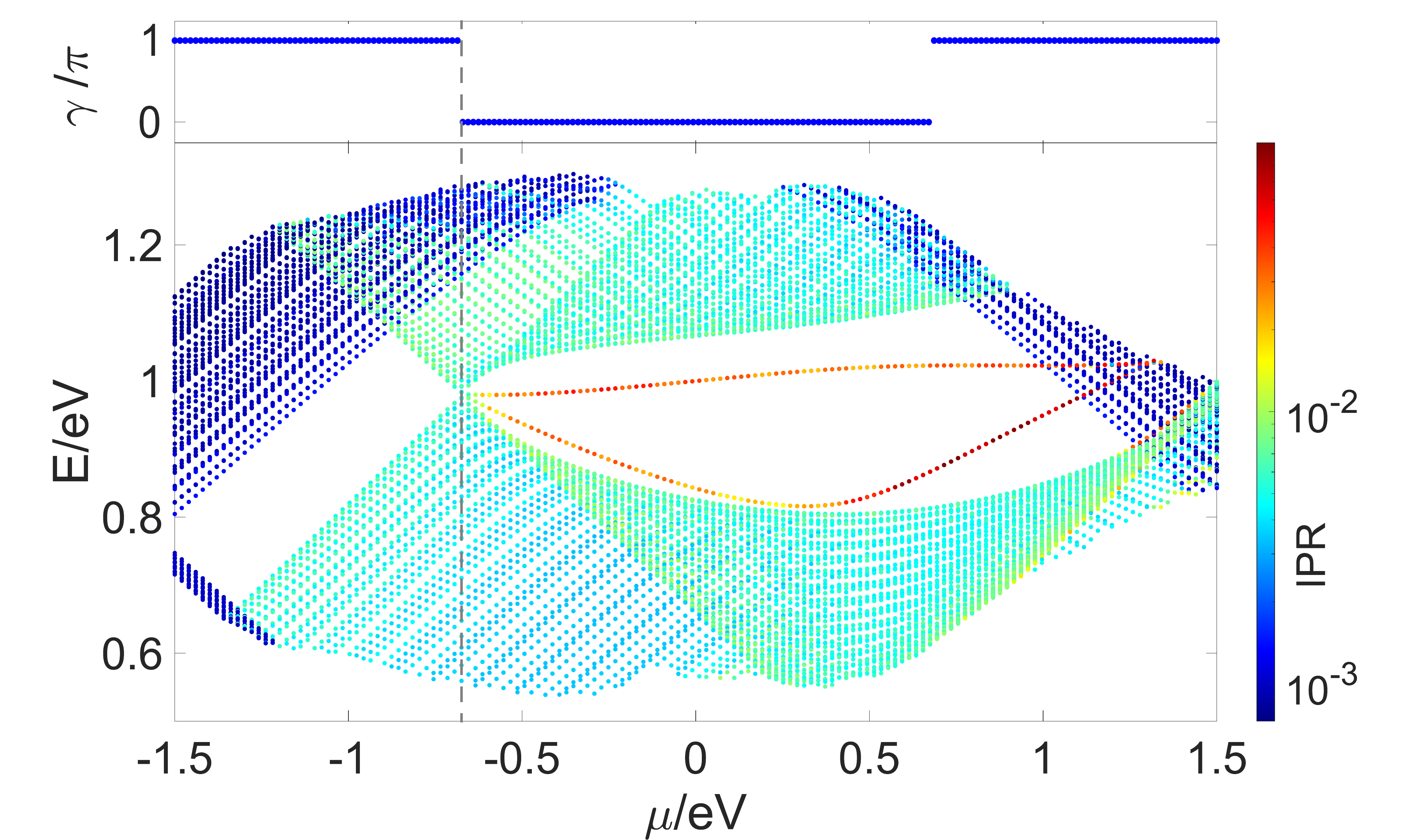}
    \caption{Energy spectrum and Berry phase $\gamma$ versus the layer-dependent on-site energy $\mu$, for the triangular lattice with $\nu=0.3$ eV. $200$ energy points around $E=0.92$ eV are taken in our numerics.The side length of the lattice is chosen to be $70$ M atoms.}
    \label{Energy_vs_mu_tri}
\end{figure}

%with different energy. Distribution in real space of corner states is shown in Figure.\ref{fig3_tri}(c)-(h), inwhich states 1, 2 and 3  are three-fold degenerate states with higher energy, states 4, 5 and 6 are three-fold degenerate states with lower energy in Figure.\ref{fig3_tri}(c)-(e), (f)-(h), respectively.

In Fig. \ref{Energy_vs_mu_tri}, we illustrate the eigenenergies around the energy gap as a function of the layer-dependent on-site energy $\mu$, which are marked by different colors according to the inverse participation ratio (IPR) of their eigenstates, defined as 
$${\rm IPR}(n)=\sum_i\left[\sum_{\alpha} \left(|\psi^{\rm 1st}_{i,\alpha,n}|^2+|\psi^{\rm 2nd}_{i,\alpha,n}|^2\right) \right]^2.$$
%$${\rm IPR}(n)=\sum_i\left[\left(\sum_{\alpha} |\psi^{\rm 1st}_{i,\alpha,n}|^2\right)^2+\left(\sum_{\alpha} |\psi^{\rm 2nd}_{i,\alpha,n}|^2\right)^2\right]$$
As can be seen from the figure, by turning on the layer-dependent on-site energy $\mu$, the energy gap closes between 1D edge states at $\mu\approx-0.67$ eV, and corner states disappear when further decreasing the (negative) value of $\mu$.
Such an observation matches a jump of the Berry phase $\gamma$ from $0$ to $\pi$, 
yet with two unconventoinal properties seemingly contradictory to conventional bulk-boundary correspondence of topological phases:
(i) corner states emerge with a trivial Berry phase ($\gamma=0$), and disappear with a nontrivial one ($\gamma=\pi$);
and (ii) the same edge-gap closing does not occur at the other topological transition point of $\mu\approx 0.67$ eV.

To understand these enigmatic behaviors, we note that the Berry phase is calculated under the nanoribbon geometry with both M-X and X-M edges, and the ``trivial" phase with $\gamma=0$ is in fact nontrivial with single-band Berry phases of $\pm \pi$ for the two edge bands, as discussed for Fig. \ref{fig:Evsmu} and in Appendix \ref{Berry phase}.
The triangular lattice we consider in Fig. \ref{fig3_tri} possesses only M-X edges, therefore it only inherits ``half" of the topological properties of a nanoribbon geometry, which is trivialized when $\mu\lesssim -0.67$ eV. 
Alternatively, a triangular lattice with X-M edges shall inherit the other ``half" of the topological properties,
possessing a spectrum symmetric to Fig. \ref{Energy_vs_mu_tri} regarding $\mu=0$, with a topological phase transition at $\mu\approx 0.67$ eV (no shown).
As a side note, a large amplitude of $\mu$ will shift different bulk bands of the two layers and mix them in energy, thus all corner states will eventually merge into the bulk bands even without a topological transition, as seen in Fig. \ref{Energy_vs_mu_tri} with $\mu\gtrsim 1$ eV.

\subsection{Hexagonal structure}\label{hexagonal structure}
\begin{figure}
    \includegraphics[width=\linewidth]{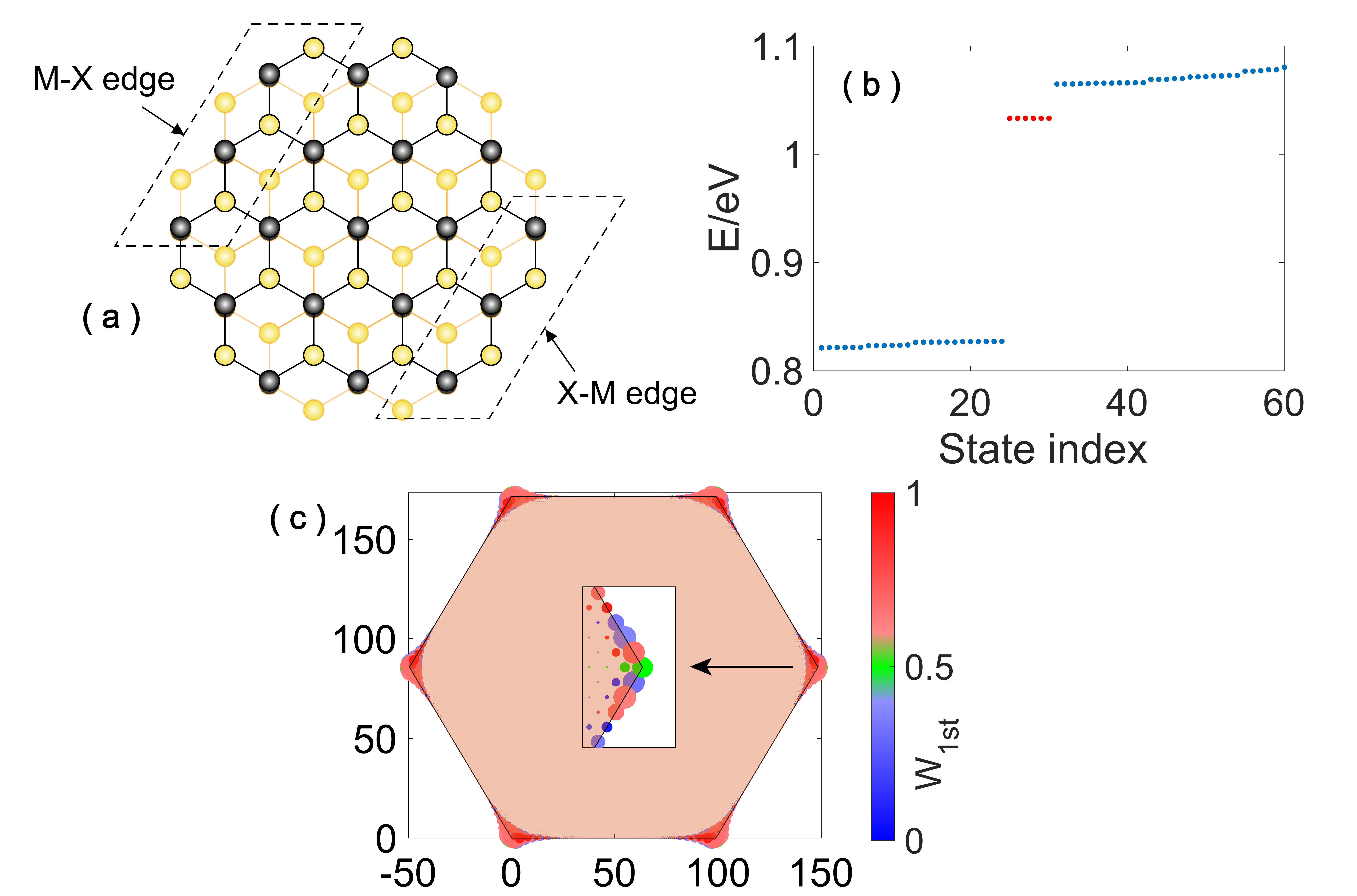}
    \caption{Lattice structure and corner states in AB-stacked bilayer TMDs with a hexagonal structure. 
    (a) Top view of the hexagonal lattice.
    (b) $60$ Eigenenergies close to $E=0.95$ eV, with a set of six-fold degenerate corner states (red) within the energy gap.
    (c) distribution of the six degenerate corner states.
    The size of each point is proportional to the summed distribution $\rho_{i}$ at site $i$, and the colormap displays the value of $W_{i}^{\rm 1st}$.
    The system's size is chosen to have $100$ M atoms along each edge. Other parameters are $\nu=0.3$ eV and $\mu=0$.
    }\label{fig5_hex}
\end{figure}

Next we consider a hexagonal structure of the AB-stacked bilayer TMDs, as sketched in Fig.\ref{fig5_hex}(a).
Unlike the triangular lattice, a hexagonal lattice have adjacent M-X and X-M edges, and satisfies both the inversion symmetry of $\hat{I}=\sigma_x \hat{\bar{R}}$ and the 3-fold rotation symmetry of $\hat{C}_3$. Consequently, corner states in the hexagonal lattice shall be six-fold degenerate, and distributes evenly on the two layers, as verified by our numerical results in Fig. \ref{fig5_hex}(b) and (c).

\begin{figure}
    \includegraphics[width=\linewidth]{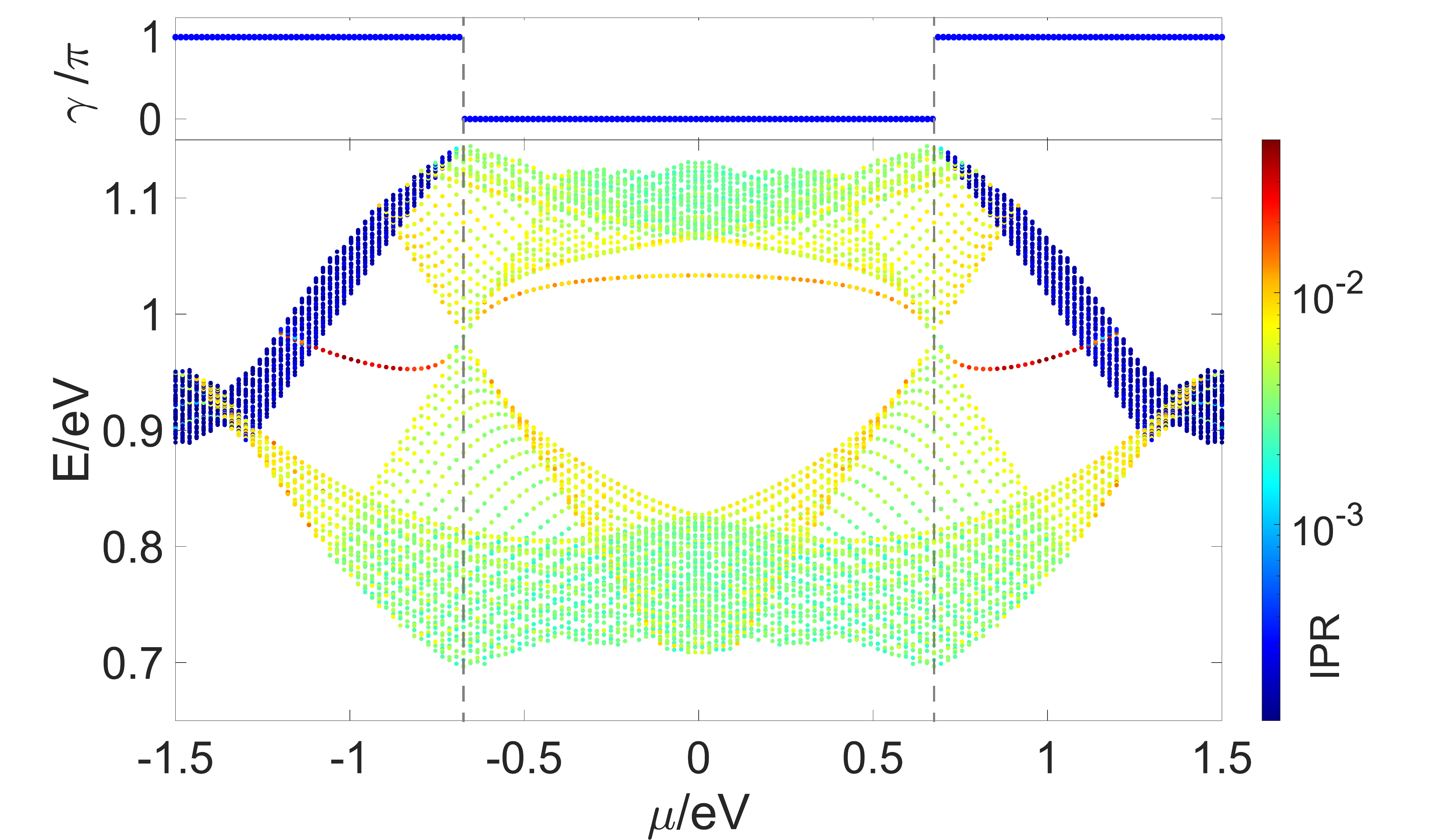}
    \caption{Energy spectrum and Berry phase $\gamma$ versus the layer-dependent on-site energy $\mu$, for the hexagonal lattice with $\nu=0.3$ eV. $200$ energy points around $E=0.92$ eV are taken.The side length of the lattice is chosen to be $70$ M atoms.}
    \label{Energy_vs_mu_hex}
\end{figure}

Due to the different boundary terminations, corner states of the hexagonal lattice behave rather differently across the topological phase transition characterized by the Berry phase $\gamma$. As demonstrated in Fig.~\ref{Energy_vs_mu_hex}, the six-fold degenerate corner states survive in all the three parameter regions separated by the gap closing at $\mu\approx\pm 0.67$ eV (except that they merge into bulk bands for $\mu\gtrsim 1$ eV).
This is because under a hexagonal geometry, each corner connects two distinct M-X and a X-M edges, and at least one of them is governed by nontrivial topology in the three parameter regions, resulting in a corner state at each of their joint corners.

\subsection{Parallelogram structure\label{Parallelogram}}
\begin{figure}
    \includegraphics[width=\linewidth]{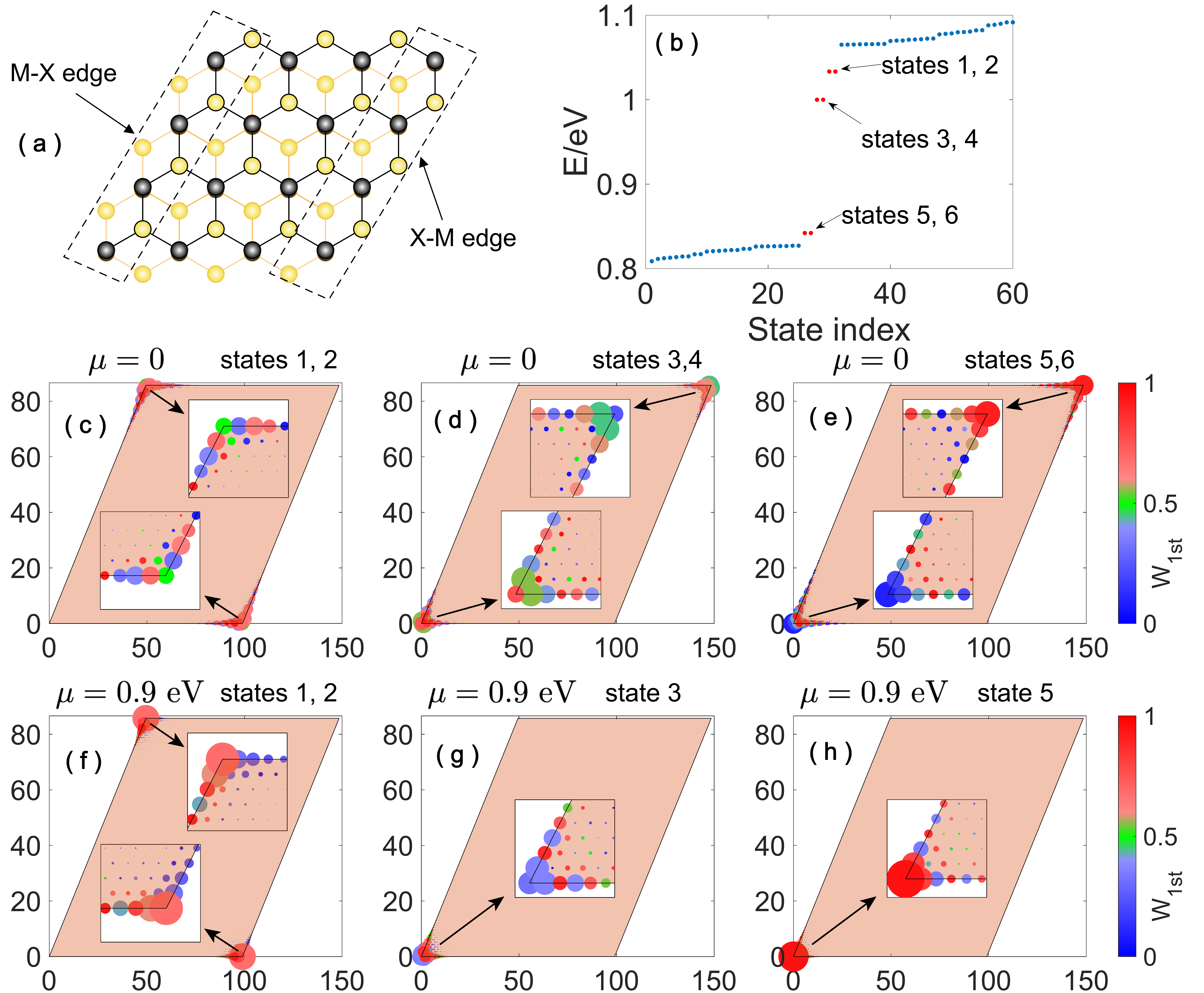}
    \caption{
    Lattice structure and corner states in AB-stacked bilayer TMDs with a parallelogram structure. 
    (a) Top view of the parallelogram lattice.
    (b) $60$ Eigenenergies close to $E=0.95$ eV at $\mu=0$, with three sets of two-fold degenerate corner states (red) within the energy gap.
    (c) to (e) distribution of the three sets of degenerate corner states respectively, with $\mu=0$.
    The size of each point is proportional to the summed distribution $\rho_{i}$ at site $i$, and the colormap displays the value of $W_{i}^{\rm 1st}$.
    The system's size is chosen to have $N_x=N_y=100$, with $N_x$ and $N_y$ the number of M atoms along $x$ and $y$ directions respectively.
    (f) distribution of the degenerate corner states (1,2) with $\mu=0.9$.
    (g) and (h) distribution of a single non-degenerate corner state with $\mu=0.9$. These two states are marked as states 3 and 5, as they correspond to states (3,4) and (5,6) in a different topological phase with $|\mu|\lesssim 0.67$ eV (see Fig. \ref{Energy_vs_mu_para}).
    $\nu=0.3$ eV is chosen for all panels.}
    \label{fig7_para}
\end{figure}

From the aspect of boundary terminations, a parallelogram structure can be viewed as a mixture of triangular and hexagonal ones, as it support both types of corners connecting either the same or two different M-X and X-M edges, as sketched in Fig. \ref{fig7_para}(a).
Consequently, corner states in a parallelogram lattice are expected to behave differently on different types of corners, 
which is also suggested by the lack of the $C_3$ rotation symmetry in such a geometry.
%Shearing along ${\rm R}_1$ and ${\rm R}_2$, AB-stacked TMDs with parallelogram structure is obtained as Figure.\ref{fig7_para}(a). The four sides of parallelogram structure have both ``M-X edge'' and ``X-M edge'', respectively. Two angles adjacent each other consist of homogeneous and heterogeneous ``M-X edge'' or ``X-M edge''. 
%Symmetric operations' set is $\{  \hat{E},\hat{M}_1,\hat{I} \}$ when $N_x=N_y$, $\{  \hat{E},\hat{I} \}$ when $N_x\ne N_y$, where $N_x,N_y$ are the numbers of M atoms along x-direction and y-direction in Figure.\ref{fig7_para}(a), respectively.

\begin{figure}
    \includegraphics[width=\linewidth]{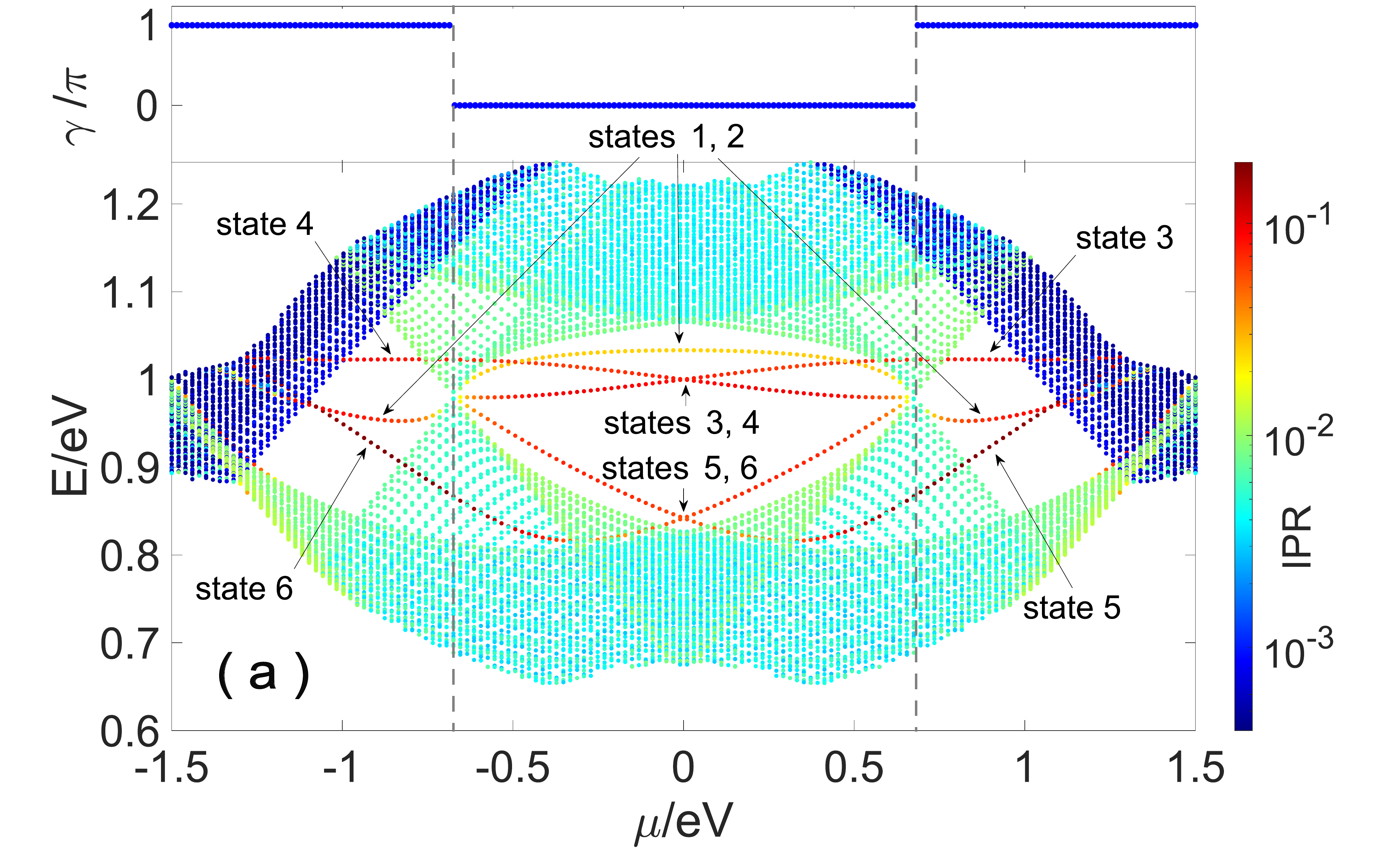}
    \caption{
    Energy spectrum and Berry phase $\gamma$ versus the layer-dependent on-site energy $\mu$, for the parallelogram lattice with $\nu=0.3$ eV. $200$ energy points around $E=0.92$ eV are taken.The size of the lattice is chosen to be $N_x=N_y=70$.}
    \label{Energy_vs_mu_para}
\end{figure}

The energy spectrum around the zigzag edge gap of AB-stacked bilayer TMDs in a parallelogram geometry is shown in Fig.\ref{fig7_para}(b),
where three sets of of two-fold degenerate corner states are found inside the gap.
Their distribution in real space is illustrated in Fig. \ref{fig7_para}(c) to (e),
where one set of corner states (labeled as states 1 and 2) occupy the top-left and bottom-right corners, 
in analogous to the corner states in the hexagonal lattice;
and the rest two (labeled as states 3 to 6) occupy the other two corners, in analogous to the corner states in the triangular lattice.
Remarkably, due to the presence of the inversion symmetry ($\hat{I}=\hat{R}\sigma_x$), each set of corner states now exhibits symmetric distribution on the two layers, in contrast to the case of the inversion-asymmetric triangular lattice.

Due to their different boundary terminations, these corner states behave rather differently upon turning on a nonzero $\mu$.
As demonstrated in Fig. \ref{Energy_vs_mu_para}, the hexagonal-analogous corner states (1,2) remain degenerate and exist for all values of  $\mu$, as they both originate from the same type of joint corners between M-X and X-M edges. 
In contrast, the two triangular-analogous corner states of a degenerate pair [(3, 4) or (5, 6)] originate from corners of different types of edges. That is, the bottom-left corner connect two M-X edges, and the top-right corner connect two X-M edges.
As discussed previously for the triangular lattice, these two types of corners react to $\mu$ differently (but symmetric about $\mu=0$),
therefore their degeneracy is lifted by a nonzero $\mu$.
Further increasing the amplitude of $\mu$, a topological phase transition occurs for M-X (X-M) edge when $\mu <0$ ($\mu >0$),
and one corner state of each triangular-analogous pair disappear when $|\mu|\gtrsim 0.67$ eV.
The disappearing of two corner states at the same corner is in consistence with our results and analysis for the triangular lattice,
which hosts three $C_3$-rotation-symmetric corners connecting the same type of edges.

\subsection{Symmetry-protected degeneracy of corner states}
\begin{figure}
    \includegraphics[width=\linewidth]{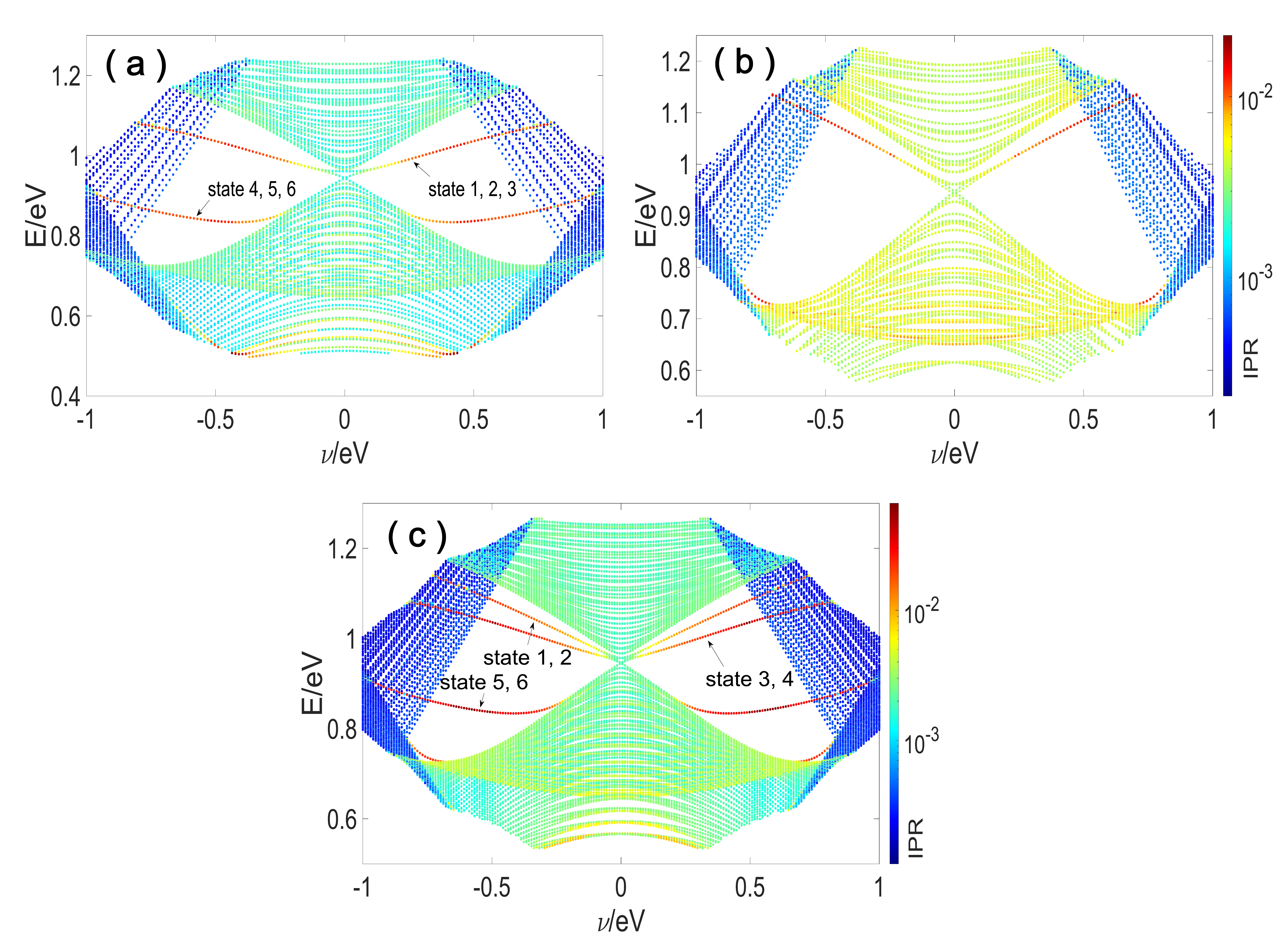}
    \caption{Energy spectrum at $\mu=0$ versus interlayer hopping strength $\nu$ for 
    (a) the triangular structure with $200$ points around $0.87$ eV,    
    (b) the hexagonal structure with $300$ points around $0.9$ eV, and 
    (c) the parallelogram structure with $200$ points around $0.9$ eV. 
    }\label{Energy_vs_nu}
\end{figure}

The above observations and analysis imply that in the parallelogram lattice, the two-fold degeneracy of triangular-analogous corner states is protected by the inversion symmetry, and the three-fold degeneracy of corner states in the triangular lattice is protected by the $C_3$ rotation symmetry, rather than topological properties related to the Berry phase. 
Indeed, as shown in Fig. \ref{Energy_vs_nu} that displays the energy spectrum versus $\nu$ for the three different geometries,
some of these corner states [states 4, 5, and 6 in (a) and states 5 and 6 in (c)] emerge from bulk bands when $\nu$ exceeds a nonzero value, without involving gap closing between different bands.
On the other hand, corner states in the hexagonal lattice, and those hexagonal-analogous ones in parallelogram lattice,
arise from connection of heterogeneous edges and are always degenerate, manifesting the topological protection corresponding to the Berry phase.

%Finally, we note that the distribution on the two layers also become asymmetric for both triangular-analogous and hexagonal-analogous corner states [as shown in Fig. \ref{fig7_para}(f) to (h)], as the inversion symmetry is broken by $\mu$.

\section{Summary and discussion\label{Conclusion}}

We have unveiled a class of extrinsic higher-order topological phases induced by interlayer couplings in AB-stacked bilayer TMDs, which host corner states insensitive to crystal symmetries of the materials. 
Explicitly, the zigzag edge states of these materials cross each other in their eigenenergy, thus a band inversion between them can be induced by nonzero interlayer couplings, leading to gapped edge bands and the emergence of in-gap corner states.
With exhaustive investigation into the system with different triangular, hexagonal, and parallelogram geometries, we uncover two types of corner states corresponding to different corner terminations between the M-X and X-M zigzag edges for the bilayer structure.
The topological nature of these corner states is justified by a multiband Berry phase defined for the system in a nanoribbon geometry,
which reflects the overall topological properties of both types of zigzag edges.
Topological phase transitions of the system induced by a on-site energy detuning $\mu$ of the two layers are studied accordingly.
We find that the M-X and X-M edges can be trivialized only for negative and positive $\mu$ respectively, leading to distinct behaviors for the two types of corner states when a topological phase transition occurs.
Our results show that the higher-order corner states in bilayer TMDs are highly tunable due to the abundant bilayer structures and layer-dependent physical effects (such as a perpendicular electric field), and thus hold great promise for quantum applications.

Throughout our study, we have focused only on the zigzag edges, as they support gapless edge states and thus an edge band inversion can be induced by weak interlayer couplings.
In Appendix \ref{app:armchair}, we have further demonstrated numerical results of the AB-stacked bilayers TMDs in a square lattice, which support both zigzag and armchair edges. 
We find that corner states in square lattices may be attributed to either zigzag- or armchair-edge band inversion, yet the later occurs only with a relatively large interlayer hopping strength, due to the large armchair-edge band gap of monolayer TMDs.
%It has been shown that armchair edges of monolayer TMDs already support higher-order corner states, which can co-exist with the ones here originating from nontrivial topology of zigzag edges. 
%However, we note that the three-orbital tight-binding model we employ here is most  suitable for describing physics of $\pm K$ valleys and zigzag edges~\cite{liu_three-band_2013}, but may be less accurate for armchair edge states which involve $\Gamma$ and $M$ points.
%Their explicit behaviors and interplay with corner states of zigzag edges in bilayer structures are beyond the scope of this work and await further investigation.

\section{Acknowledgements}
This work is supported by the Guangdong Project (Grant No. 2021QN02X073).

\appendix
\section{Single-band Berry phases \label{Berry phase}}
In this appendix we discuss properties of single-band Berry phase $\gamma_n$ for the zigzag edges of AB-stacked bilayer TMDs.
It follows a similar definition as in Eq. \eqref{eq_BP} in the main text,
\begin{equation}
    \gamma_n=- i \sum_l \log  U_n(k_l),
\end{equation}
with $U_n(k_l)=\langle\psi_n(k_l)|\psi_n(k_{l+1}\rangle$ and $\psi_n(k_l)\rangle$ the Bloch wavefunction of the $n$-th band at the discrete crystal momentum $k_l$. Here we consider the two degenerate edge bands belong the energy gap in Fig. \ref{fig1_model}(b), labeled as $\psi_{rm MX}$ and $\psi_{\rm XM}$ respectively regarding their occupied edges.
Due to the presence of inversion symmetry, these two edge bands are symmetric to each other between $k$ and $-k$, i.e. $$\hat{I}_y|\psi_{\rm MX}(k)\rangle=|\psi_{\rm XM}(-k)\rangle,$$ with $\hat{I}_y=\hat{M}_y\sigma_x$ the inversion operation in $y-z$ plane (the same as $\hat{M}'_y$ discussed in Sec. \ref{sec_sym}).
In addition, the model also satisfies the time reversal symmetry $\hat{H}(k)=\hat{H}^*(-k)$, or $|\psi(k)\rangle=|\psi^*(-k)\rangle$ for eigenstates.
Thus the link variable $U_{n}(k)$ for these two bands satisfies
\begin{equation}
    \begin{aligned}
        U_{\rm MX}(k)&=\langle{\psi_{\rm MX}(k_l)}|{\psi_{\rm MX}(k_{l+1})}\rangle\\
        &=\langle{\psi_{\rm XM}(-k_l)}|{\psi_{\rm XM}(-k_{l+1})}\rangle\\
        &=\langle{\psi_{\rm XM}(k_l)^{*}}|{\psi_{\rm XM}(k_{l+1})^{*}}\rangle\\
        &=\langle{\psi_{\rm XM}(k_l)}|{\psi_{\rm XM}(k_{l+1})}\rangle^{*}\\
        &=U_{\rm XM}(k)^{*}
    \end{aligned}
\end{equation}
which leads to 
$\gamma_{\rm MX}=-\gamma_{\rm XM}$. Therefore the Berry phase of two boundary states vanish at on-site energy $\mu=0$.

Although these edge-band Berry phases are obtained for $\mu=0$, their symmetric behavior is expected to hold even when the inversion symmetry is broken by a nonzero $\mu$, unless a topological phase transition occurs for one of the two edge bands.
However, one of the two edge-band Berry phases may be ill-defined when the amplitude of $\mu$ increases, since the corresponding edge band may partially merge into the valence bulk bands [e.g. see Fig. \ref{fig:Evsmu}(b) and (c) in the main text]. 
Therefore we have only considered the multi-band Berry phase $\gamma$ in the main text.

\section{Corner states of armchair boundaries}\label{app:armchair}
%armchair本身就有gapped 边界态，并存在角态。增加层间耦合，原则上可以产生类似zigzag的能带反转，从而产生新的角态。这两种边界的角态理论上可以在方形几何共存。但注意到，三带模型对armchair边界态的描述误差大一些，armchiar的gap会比较大（文献？），因此这里的结果与实际情况可能具有较大偏差。

\begin{figure}
    \includegraphics[width=\linewidth]{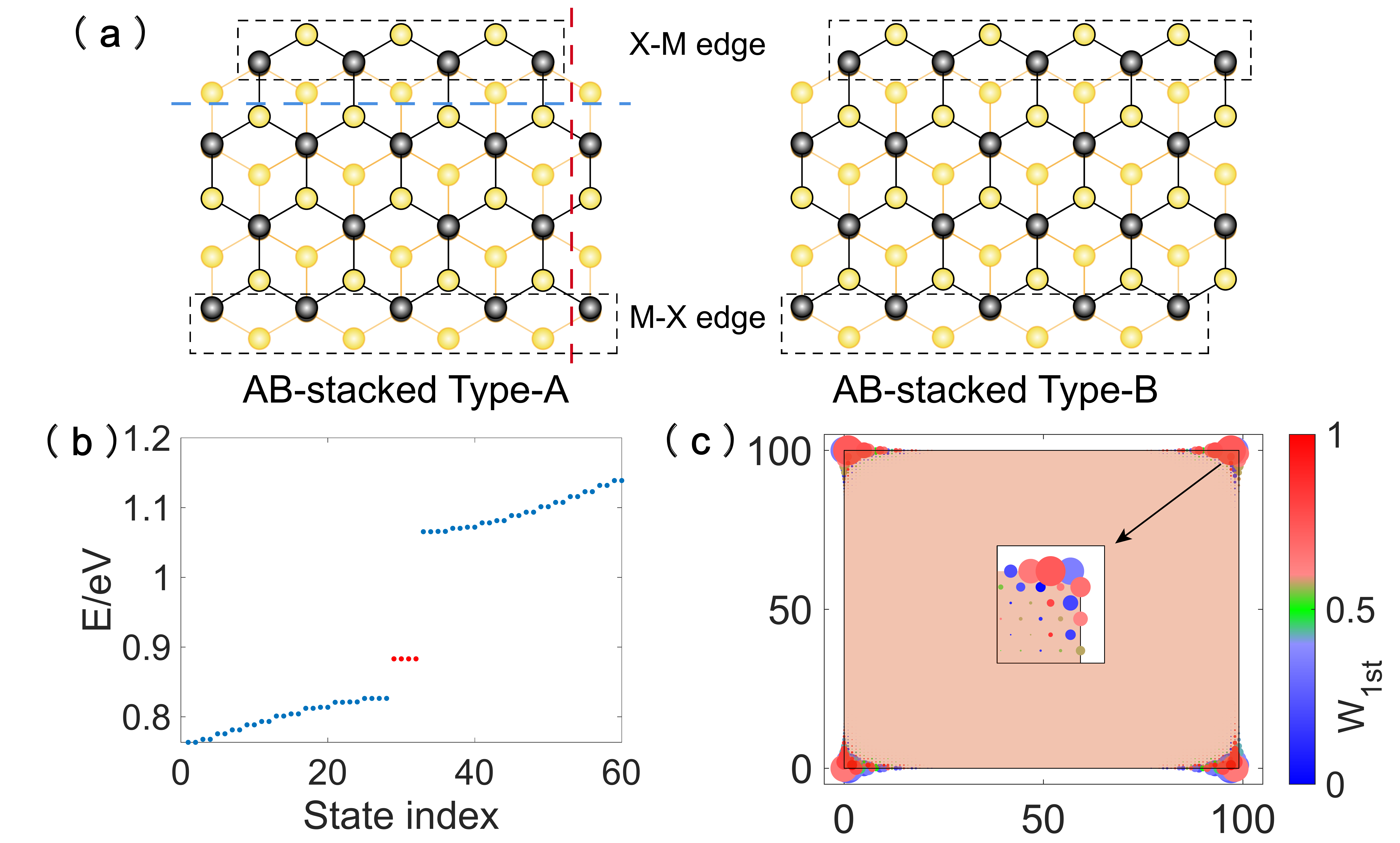}
    \caption{Lattice structure and corner states in AB-stacked bilayer TMDs with square structures.
      (a) Top view of different types square structures. A type-B lattice can be obtained from a type-A one by shearing along the red dash line.
      Shearing along the blue dash line changes the parity of $N_y$.
      (b) $60$ Eigenenergies close to $E=0.95$ eV at $\mu=0$ for the type-A square lattice with an odd $N_y$, with four degenerate corner states (red) within the energy gap.
      (c) distribution of the the degenerate corner states, where the size of each point is proportional to the summed distribution $\rho_{i}$ at site $i$, and the colormap displays $W_{i}^{\rm 1st}$, the weight of the first-layer occupation on each site.
      The system's size is chosen to be $N_x=100$ and $N_y=101$.
      Other parameters are $\mu=0$ and $\nu=0.3$ eV.
      }\label{square} 
\end{figure}

%\subsection{Crystal structure and crystal symmetry of square structure}
In this appendix we discussed AB-stacked TMDs with square structure, which host both armchair and zigzag edges.
%next we will discuss AB-stacked TMDs with square strucutre which host armchair and two types of zigzag boundaries:``M-X edge'' and ``X-M edge''.
Taking into account two types of M-X and X-M zigzag edges,
we obtain four types of different nano-flakes, namely type-A and type-B denoted in Fig. \ref{square}(a), with either an even or an odd $N_y$, the number of M atoms along $y$ direction.
Their corresponding symmetries are indicated in Table \ref{symmetry with square}.

\begin{table}
  \caption{Symmetry table for different types of AB-stacked bilayer TMDs with square structure. $N_y$ is the number of M atoms along $y$-direction.}
  \begin{tabular}{lcc}
    \hline \hline
     &  Type A &  Type B\\
    \hline
    $N_y$ is odd $(N_y>1)$ & $\left\{\hat{E}, \hat{M}_x, \hat{I}, \hat{M}'_y\right\}$ & $\left\{\hat{E},\hat{M}'_y\right\}$ \\
    $N_y$ is even & $\left\{\hat{E}, \hat{M}_x \right\}$ & $\left\{\hat{E},\hat{I}\right\}$\\
    \hline \hline
 \end{tabular}
 \label{symmetry with square}
\end{table}

\begin{figure}
    \includegraphics[width=\linewidth]{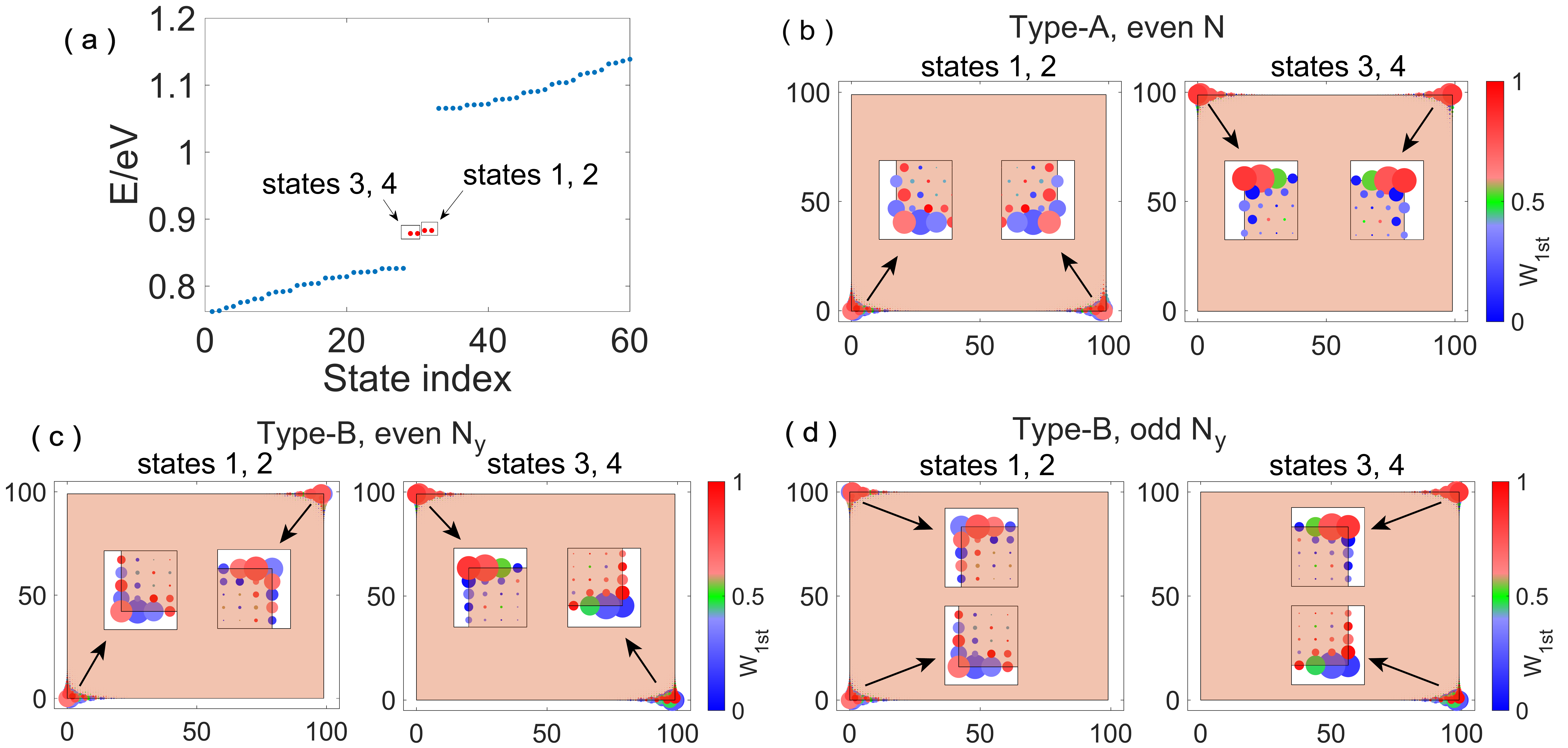}
    \caption{
     Spectrum and corner states for different types of square lattices.
     (a) $60$ Eigenenergies close to $E=0.95$ eV at $\mu=0$ for the type-A square lattice with an even $N_y$, with two sets of two-fold degenerate corner states (red) within the energy gap.
     Spectra for type-B square lattices (for both even and odd $N_y$) are qualitatively the same as (a).
      (b) to (d) distribution of the the degenerate corner states
      for type-A with even $N_y$ ($N_x=100, N_y=100$), type-B with even $N_y$ ($N_x=100, N_y=100$), and type-B with odd $N_y$ ($N_x=100, N_y=101$), respectively.
      The size of each point is proportional to the summed distribution $\rho_{i}$ at site $i$, and the colormap displays $W_{i}^{\rm 1st}$, the weight of the first-layer occupation on each site.
      Other parameters are $\mu=0$ and $\nu=0.3$ eV.}
    \label{square2}
\end{figure}

As seen in Fig. \ref{square}(b) and (c), a type-A square lattice with an odd $N_y$ has four degenerate corner states, distributing evenly on the four corners.
This is because it is the most symmetric case in the four types of square lattices, satisfying both the inversion symmetry $\hat{I}$ and mirror symmetries $\hat{M}_{x,y}$.
On the other hand, the rest three types of square lattices are less symmetric, and the corner states split into two sets of two-fold degenerate pair, as demonstrated in Fig. \ref{square2}(a) for the example of a type-A square lattice with an even $N_y$.
Their spatial distributions are illustrated in Fig. \ref{square2}(b) to (d), which reflects the $\hat{M}_x$, $\hat{I}$, and $\hat{M}'_y$ symmetries of the corresponding square structures respectively.

\begin{figure}
    \includegraphics[width=\linewidth]{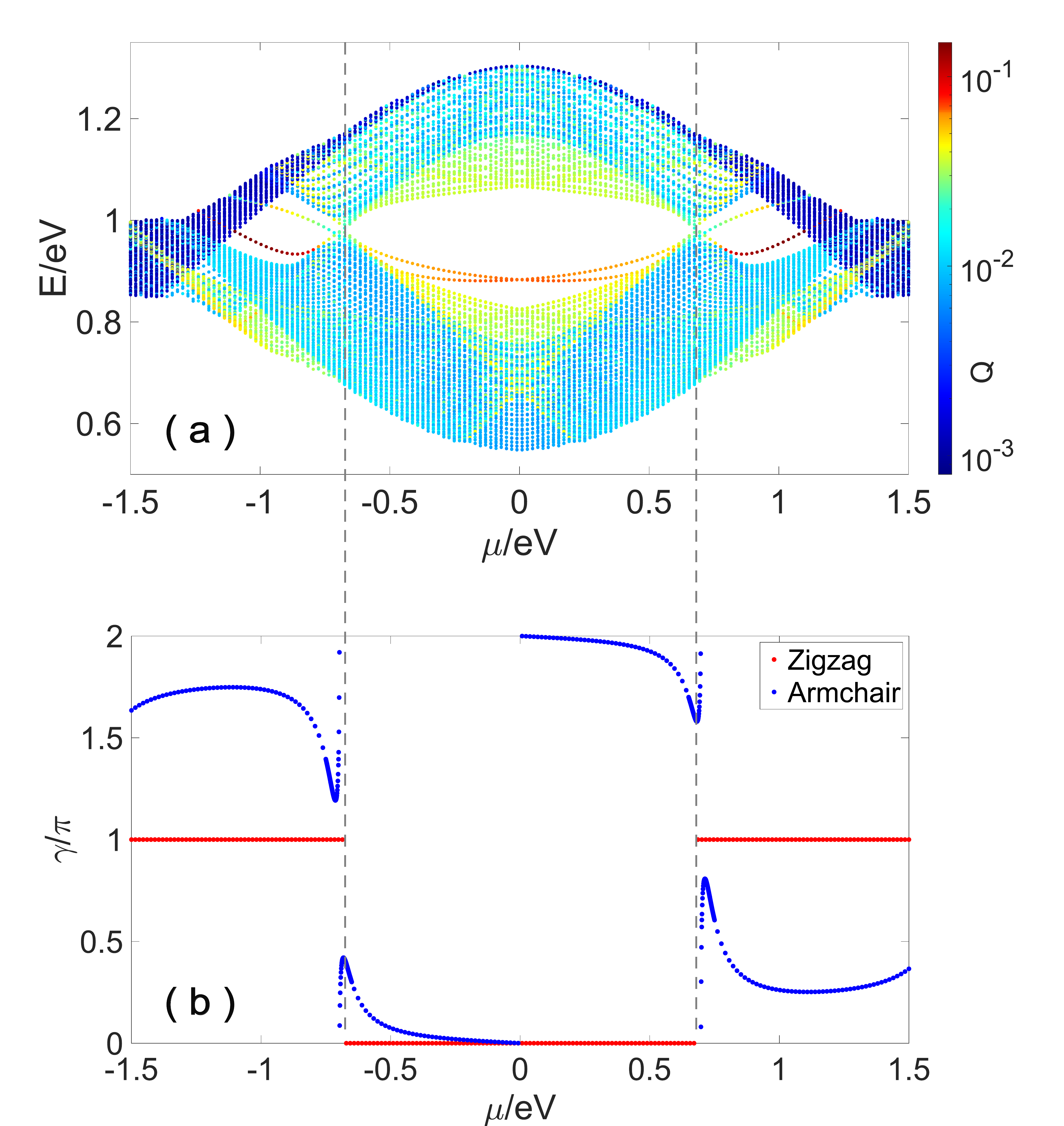}
    \caption{(a)Eigenenergy versus on-site energy
    in AB-stacked bilayer TMDs with Type-A structure and odd $N_y$. 
    200 points around $0.92$ eV are taken. The size of the lattice $N_x=71,N_y=71$. The interlayer coupling strength is chosen to be  $\nu=0.3$ eV.
    (b) Multiband Berry phases for all eigenstates below the edge gap, for zigzag (red) and armchair (blue) nanoribbons respectively.}\label{Berry_square}
\end{figure}

In Fig. \ref{Berry_square}(a), we demonstrate the energy spectrum as a function of $\mu$, for type-A square lattice with odd $N_y$, namely with both X-M and M-X zigzag edges.
Each state is marked by a quantity $Q$ defined to distinguish armchair and zigzag edge states,
\begin{equation} \label{IPR}
    \begin{aligned}
        Q&=({\rm IPR})^{\frac{4}{5}}\cdot ({\rm IPR}_x)^{\frac{1}{5}}, \\
        {\rm IPR}&=\sum_i\left[\sum_{\alpha} \left(|\psi^{\rm 1st}_{i,\alpha,n}|^2+|\psi^{\rm 2nd}_{i,\alpha,n}|^2\right) \right]^2,\\
        {\rm IPR}_y&=\sum_{i_y}\left[\sum_{\alpha,i_x} \left(|\psi^{\rm 1st}_{i,\alpha,n}|^2+|\psi^{\rm 2nd}_{i,\alpha,n}|^2\right) \right]^2,\\    \end{aligned}
\end{equation}
where the site index $i$ is further expressed as $(i_x,i_y)$ to indicate different positions along $x$ and $y$ directions,
and ${\rm IPR}_y$ describes the localization strength along $y$ direction.
With this definition, perfectly localized zigzag and armchair edge states are expected to have $Q\approx (1/N_x)^{4/5}$ and $ 1/N_y$, as they are localized along $y$ and $x$ directions, respectively.

As shown in the figure, the four-fold degenerate corner states split into two pairs of two-fold degenerate ones, due to the breaking of inversion symmetry under a nonzero $\mu$.
The gap of both zigzag and armchair edge states is seen closes at $|\mu|\approx 0.67$ eV, the same as for the cases with zigzag edges only. 
As discussed in the main text, 
the gap closing and reopening represents a topological phase transition that trivializes one of the M-X and X-M zigzag edges,
thus only one pair of edge states can be attributed to zigzag edges when $|\mu|\gtrsim 0.67$ eV.
However, the gap closing and reopening also indicate a band inversion for the armchair edge bands, which are gapped at $\mu=0$. Thus we still observe two pairs of edge states after the transition, with one of them originating from an armchair-edge band inversion.

In Fig. \ref{Berry_square}(b), we display the Berry phases of both zigzag and armchair nanoribbons, where the later is found to be not quantized generally. Nevertheless, the armchair Berry phase is found to be $0$ at $\mu=0$, indicating that the armchair edges are topologically trivial when $\mu$ is small, i.e. within the parameter region between the two gap closing points for armchair edge states.
The rapid change of the armchair Berry phase at the gap closing points hints a transition analogous to a topological one, which reflects the origin of armchair-edge band inversion for one pair of edge states at $|\mu|\gtrsim 0.67$ eV.

%To further verified the topological transition of armchair edge bands, we calculate the multiband Berry phase for an armchair nanoribbon, as displayed in Fig. \ref{Berry_square}(b). We find that the armchair Berry phase is generally not quantized, yet it changes rapidly acoss each gap closing point, reflecting the topological nature of the transition of corner states.

%\bibliography{Reference.bib}
%

\end{document}